\documentclass[reprint,amsmath,amssymb,aps,pre]{revtex4-1}  %% z apssamp.tex / revtex4-1

\usepackage{amssymb}
\usepackage{amsmath}

\usepackage[dvips]{graphicx}

\begin{document}

\title{Hydraulic tortuosity in arbitrary porous media flow}
\author{Artur Duda}  \email{ardud@ift.uni.wroc.pl}
\author{Zbigniew Koza}
\author{Maciej Matyka}
\affiliation{Institute of Theoretical Physics, University of Wroc{\l}aw, pl. M. Borna 9, 50-204 Wroc{\l}aw, Poland, }

\date{\today}% It is always \today, today,

\begin{abstract}
Tortuosity ($T$) is a parameter describing an average elongation of fluid streamlines in a porous medium
as compared to free flow. In this paper several methods of calculating
this quantity from lengths of individual streamlines are compared
and their weak and strong features are discussed. An alternative method is proposed, which
enables one to calculate $T$ directly from the fluid velocity field,
without the need of determining streamlines, which greatly simplifies determination of tortuosity
in complex geometries, including those found in experiments or 3D computer models.
Numerical results obtained with this method
suggest that (a) the hydraulic tortuosity of an isotropic fibrous medium
takes on the form  $T = 1 + p\sqrt{1-\varphi}$, where $\varphi$ is the porosity and $p$ is a constant and (b)
the exponent controlling the divergence of $T$ with the system size at  percolation threshold
is related to an exponent describing the scaling of the most probable traveling length at bond percolation.
\end{abstract}

% 47.15.G- == Low-Reynolds-number (creeping) flows
% 47.56.+r == Flows through porous media
% 91.60.Np == Permeability and porosity

\maketitle

%\selectlanguage{english}

\section{Introduction}
Investigation of transport through porous media is of paramount importance
in many areas of science and engineering.
One of the main problems is to find out how the value of permeability,
which synthetically describes how the flow is retarded by the porous medium structure,
can be related to some more fundamental,
well-defined parameters determined solely by the geometry of the medium,
as such relation could be used, for example,
to fabricate materials of desired physical properties.

%According to the Darcy's law, the flux $\mathbf{q}$ of a Newtonian fluid flowing through
%a porous medium at low Reynolds numbers is proportional to the applied pressure gradient $\nabla P$,
%%
%%
%\begin{equation}
%  \label{eq:Darcy}
%    \mathbf{q} = -\frac{k}{\mu } \nabla P,
%\end{equation}
%%
%%
%where $\mu$ is the dynamic viscosity of the fluid and $k$ is the permeability of the medium \cite{Bear72}.
%Actually, for systems where $\mathbf{q}$ is experimentally found
%proportional to $\nabla P$, Eq.\ (\ref{eq:Darcy}) is used as the definition of $k$.
%Permeability, which synthetically describes how the flow is retarded by the porous medium structure,
%is thus of great importance for many fields science and technology
%and has been the subject of broad experimental and theoretical research.

One of the most well-known theories of this kind was developed by Kozeny and later modified by Carman \cite{Carman37}.
In their approach a porous medium is assumed to be equivalent to a bundle of capillaries
of equal length and constant cross-section.
These assumptions lead to the semi-empirical Kozeny-Carman formula \cite{Bear72,Carman37,Dullien92}
\begin{equation}
  \label{eq:Kozeny-Carman}
    k =  \frac{\varphi ^{3} }{\beta T^{2} S^{2} },
\end{equation}
which relates the permeability ($k$) to four structural parameters:
the porosity $\varphi$, the specific surface area $S$,
the hydraulic tortuosity $T$, and the shape factor $\beta$. In this equation $\varphi$ is a
dimensionless quantity defined as the fraction
of the porous sample that is occupied by pore space  ($0 < \varphi < 1$), $S$
equals to the ratio of the total interstitial surface area to the bulk volume,
$\beta$ is a constant characteristic for a particular type of granular material, and
$T$ is a dimensionless parameter defined as
\begin{equation}
  \label{eq:tortuosity}
    T = \frac{\left\langle \lambda \right\rangle }{L} \ge 1,
\end{equation}
where $\left\langle \lambda \right\rangle $ is the mean length of fluid particle paths
and $L$ is the straight-line distance through the medium in the direction of macroscopic flow.
Eq.\ (\ref{eq:Kozeny-Carman}) has been found to agree well with experimental results
for random packs of monodisperse granules, e.g.\ spheres or well sorted and rounded sands.
\begin{figure}[!t]
\centering
\includegraphics[height=0.6\columnwidth]{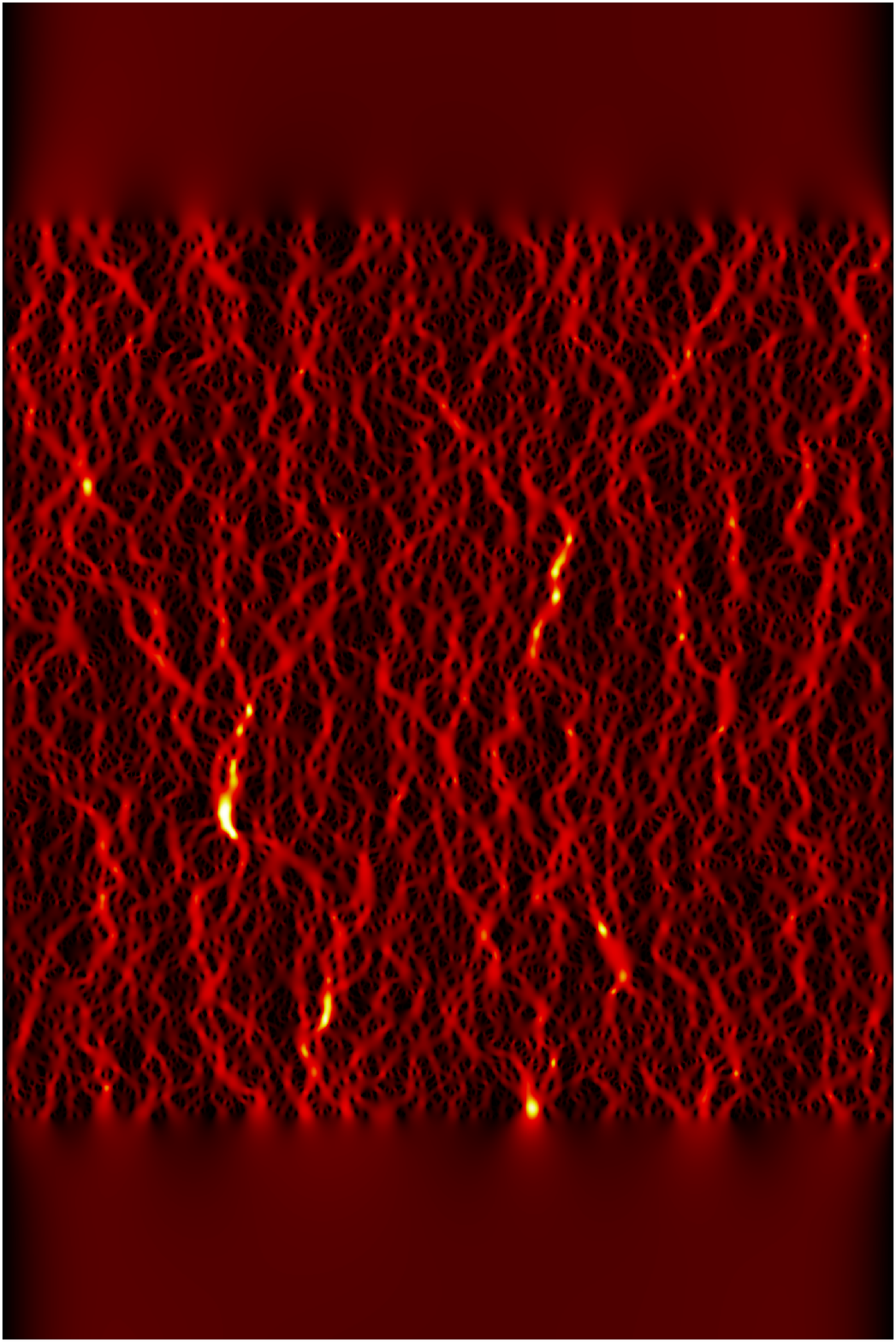}
\includegraphics[height=0.6\columnwidth]{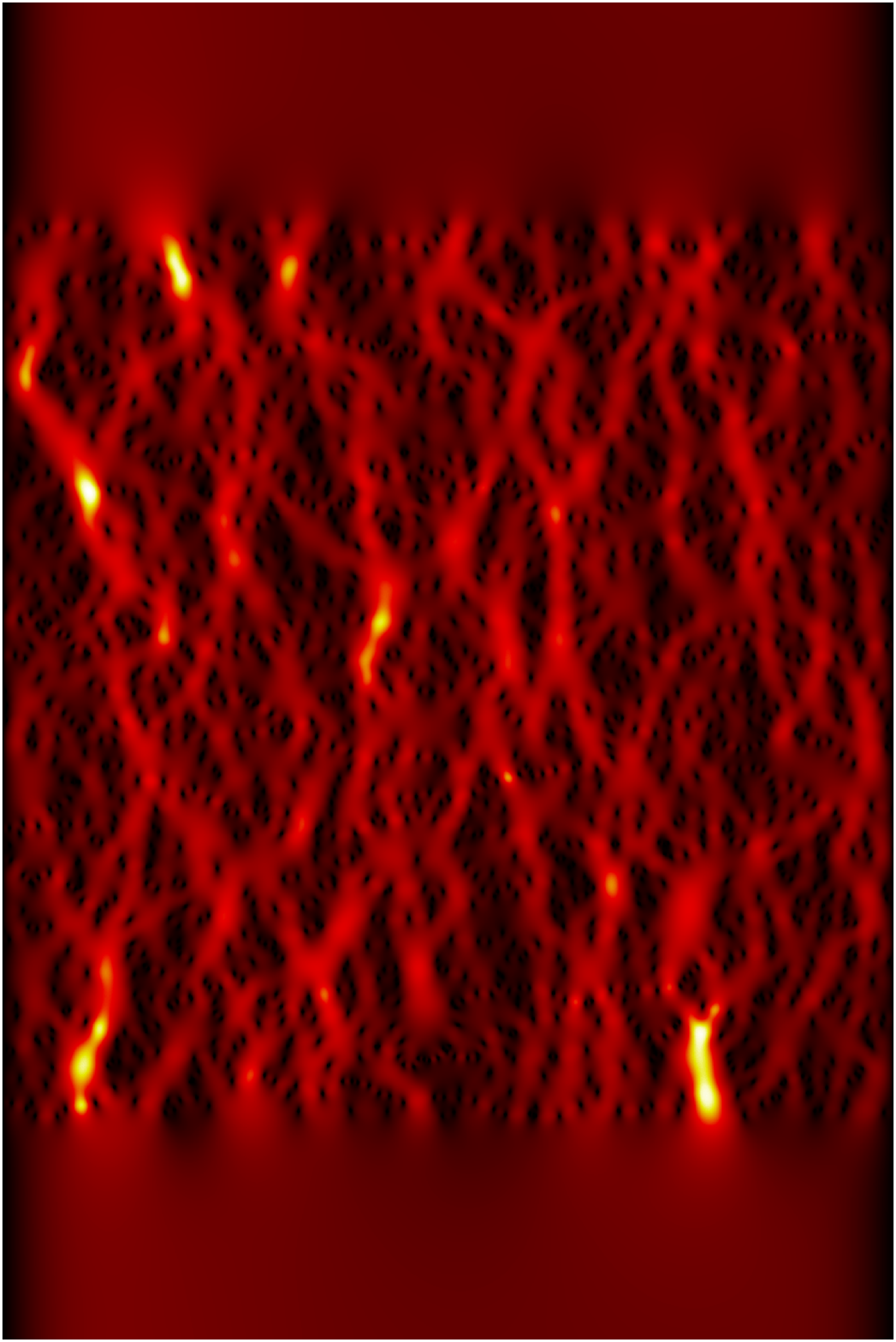}
\includegraphics[height=0.6\columnwidth]{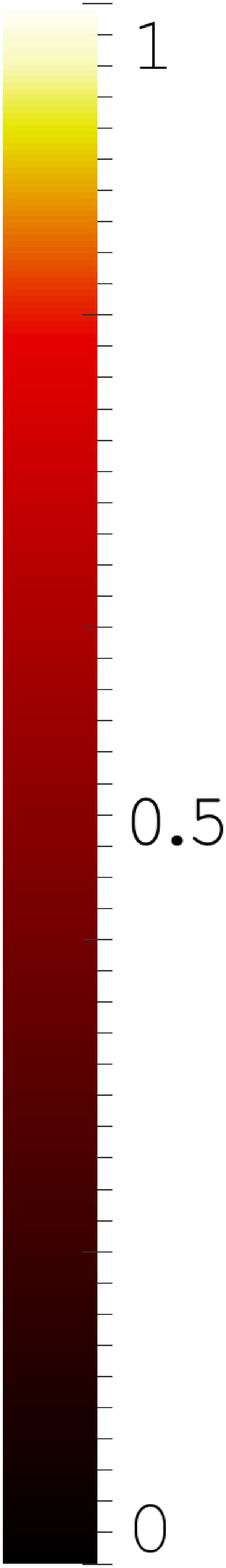}
\caption{
  \label{fig:color-at-0.99}
  (Color) Normalized velocity field ($u/u_{max}$) for two highly porous systems,
  $\varphi = 0.95$ (left) and $\varphi = 0.99$ (right).
  Lighter (darker) colors represent larger (smaller) fluid velocities.
}
\end{figure}

The Kozeny-Carman approximation of a porous medium can be used to model also other types of transport,
e.g.\ diffusion or electric current.
This observation resulted in introducing several distinctive,
experimentally measurable quantities that in the capillary
approximation can be readily linked with the tortuosity of the capillaries as given by Eq.\ (\ref{eq:tortuosity}).
In this way the term `tortuosity' has been overloaded with several essentially different meanings \cite{Clennell97},
as depending on the context it can refer not only to hydraulic, but also geometric \cite{Yu04, Yun10},
diffusional \cite{Iversen93,Garrouch01,Shen07}, electrical
\cite{Lorenz61,Tye82,Johnson82,Garrouch01}, thermal \cite{Montes07},  acoustic \cite{Johnson82},
or streamline \cite{Matyka08,Koza09} tortuosity, with no clear consensus on the relation between them.
Besides, in the literature different quantities, including $T^{-1}$, $T^{-2}$, and $T^2$
\cite{Bear72,Dullien92,Clennell97} have also been called `tortuosity'.

While the permeability in the Kozeny-Carman theory depends on four structural factors:
porosity, specific surface area, a shape factor and a hydraulic tortuosity,
until recently only the first two of them could be measured in non-trivial cases,
and only porosity could be measured relatively easily.
Carman himself attempted to estimate hydraulic tortuosity by injecting dye into a bed of glass spheres
and concluded that $T\approx\sqrt{2}$ \cite{Carman37}.
However, in practice only the product $\beta T^2$ (known as the Kozeny constant)
could be determined in non-trivial experimental setups,
which left the shape factor and hydraulic tortuosity as essentially indeterminate quantities.
This observation led several researchers to ponder whether hydraulic tortuosity really exists
as a fundamental attribute of the pore space or whether it is just a `fudge factor',
an adjustable parameter used to fit the model to the experimental data \cite{Tye82,Dullien92}.
In spite of these difficulties, diffusional and electrical tortuosities are one of the basic parameters
commonly used to characterize real porous media in such diverse areas
as medicine \cite{Sykova08,Dougherty10}, marine biology \cite{Iversen93}
or advanced materials \cite{Wilson06}, and the
hydraulic tortuosity remains a key concept of many advanced theories, e.g.\
the Effective Medium Approximation \cite{Bear72,Ahmadi2011}. Moreover, modern technology
has made it possible to determine the fluid velocity field in quite complex geometries
both experimentally \cite{Kossel05, Morad09,Onstad10} and numerically
\cite{Succi89,Koponen96,Koponen97,Matyka08,Koza09}.
This, at least in principle, enables one to determine flow streamlines
and hence renders the hydraulic tortuosity a measurable quantity.

At this point, however, there appears an unexpected problem.
A textbook recipe requires to calculate the hydraulic tortuosity as an
`average path of fluid particles through the porous medium' \cite{Bear72}
without specifying what sort of averaging is actually meant.
For example, is it to be taken over the whole volume or over a cross-section,
and in the latter case---are all cross-sections equivalent?
Should the average be weighted and how?
Ambiguity in the definition of $T$ was noticed already
by Bear \cite{Bear72}, who remarked that the average pathlines could be obtained
either by averaging the pathlines of all fluid particles passing a given cross-section of the medium
at a certain instant of time (geometrical approach) or during a given period of time (kinematical approach).
Bear himself preferred the geometrical approach, but
he never explained how his tortuosity tensor, which is a macroscopic quantity,
could be calculated from microscopic flow streamlines. Clennell \cite{Clennell97}
gave several convincing arguments in favor of the opinion that the hydraulic tortuosity should be
calculated as a kinematical average in which the pathlines are weighted with fluid fluxes.
However, until very recently a lack of precision in the definition of $T$ was not regarded as a problem,
as this quantity was considered to be too difficult to be calculated in a general case, and
most attempts in this direction concentrated on rather simple models
where the results did not depend on the averaging procedure.
Under these circumstances, when in recent years it became possible
to simulate flows numerically with unprecedented accuracy
in complex geometries  in which fluid fluxes continuously
change in sectional area, shape and orientation as well as branch and rejoin,
%so that it is impossible to reduce the system to pores connected by simple capillaries,
researchers developed their own methods of calculating $T$
\cite{Knackstedt94,Koponen96,Koponen97,Matyka08,Koza09}.
Closer inspection of these papers leads to a rather surprising conclusion
that no consensus has been reached as to the actual meaning of the numerator in
Eq.~(\ref{eq:tortuosity}), as
each research group interpreted the average in Eq.~(\ref{eq:tortuosity})
in their own, unique way.

%Closer inspection of several attempts to calculate a hydraulic tortuosity from numerical simulations
% leads to a rather surprising conclusion
%that no consensus has been reached as to the actual meaning of the enumerator in
%Eq.~(\ref{eq:tortuosity}).

The aim of this study is to propose a universal, efficient
 method of calculating hydraulic tortuosity in an arbitrary geometry.
To this end in Sec.~\ref{sec:methods} we analyze several algorithms
used so far and show their weak and strong features.  Then, in Sec.~\ref{sec:new_method}
we present a new method, which enables one to calculate $T$ without the need of determining
any streamlines, which is often an ill-conditioned numerical problem \cite{Matyka08},
especially if only approximate values of the velocity field are  available.
This is the main result of the paper. Its significance lies
in that it greatly simplifies determination of hydraulic tortuosity
in experiments or three-dimensional numerical simulations.
In Sec.~\ref{sec:Applications} we apply this method in two cases that cannot be efficiently
treated with other methods: very high porosity (fibrous medium) or very low porosity (system at percolation).
Finally, Sec.~\ref{sec:Conclusions} is devoted to conclusions.

\section{Comparison of existing methods \label{sec:methods}}

As it was already mentioned, several methods of calculating the hydraulic tortuosity in an arbitrary geometry
are available in the literature.
Most of them reduce the problem to calculating
$T$ as a weighted average of the form
\begin{equation}
  \label{eq:genarel-sum-T}
  T = \frac{1}{L} \frac{\sum_i\lambda_i w_i}{\sum_i w_i},
\end{equation}
where $i$ enumerates discrete streamlines, $\lambda_i$ is the length of the $i$-th streamline, and $w_i$ is a weight.

Knackstedt and Zhang  \cite{Knackstedt94,Zhang95} used Eq.~(\ref{eq:genarel-sum-T})
with $i$ running through the
nodes of a regular lattice on the inlet cross-section and chose the weights of the form
\begin{equation}
 \label{eq:Knackstedt}
   w_i = \frac{1}{t_i},
\end{equation}
where $t_i$ is the time in which a fluid particle moves along the $i$-th streamline.
The intention behind (\ref{eq:Knackstedt}) was to weight each streamline length
proportionally to the overall volumetric flow associated with this streamline.
Since the streamlines sample the inlet plane uniformly, the weights
satisfying this condition should be proportional to $(v_{i})_x^\mathrm{in}$,
the components of the fluid velocities parallel to the macroscopic fluid flow direction
(here assumed to be directed along the $x$ axis) and measured at the points where
the $i$-th streamline cuts the inlet plane.
Note that the weights defined by (\ref{eq:Knackstedt}) might be equivalently written
as $w_i = \bar{v}_i/\lambda_i$, with
$\bar{v}_i$ being the average fluid velocity along the $i$-th streamline.
However, apart from some trivial geometries (e.g.\ straight capillaries of equal length),
fluid velocity along a typical streamline in a complex geometry, for example in a granular porous medium,
can vary by several orders of magnitude \cite{Matyka08} and hence there is no
connection between $\bar{v}_i/\lambda_i$ and $(v_{i})_x^\mathrm{in}$.
For this reason it is not clear what the quantity calculated by Knackstedt and Zhang
has to do with the actual hydraulic tortuosity.

Koponen \textit{et al.}\ \cite{Koponen96} introduced two families of tortuosities
$T_n^\mathrm{S}$, and $T_n^\mathrm{V}$, $n \in \mathbb{Z}$, defined through
\begin{equation}
  \label{eq:Koponen-tortuosity-integralS}
    (T_n^\mathrm{S})^n = \frac{\int_{A} {\widetilde\lambda^n(\mathbf{r}) v(\mathbf{r}) }\, d^2\mathbf{r} }%
    {\int_{A} v(\mathbf{r}) \, d^2\mathbf{r}},
\end{equation}
and
\begin{equation}
  \label{eq:Koponen-tortuosity-integralV}
    (T_n^\mathrm{V})^n = \frac{\int_{V} {\widetilde\lambda^n(\mathbf{r}) v(\mathbf{r}) }\, d^3\mathbf{r} }%
    {\int_{V} v(\mathbf{r}) \, d^3\mathbf{r}},
\end{equation}
where $A$ is an arbitrary cross-section perpendicular to the macroscopic fluid flow direction,
$V$ is the volume of the sample,
$\widetilde\lambda(\mathbf{r})  = \lambda(\mathbf{r})/L$ is
the tortuosity of the flow line passing through a point $\mathbf{r}$,
$v(\mathbf{r}) = |\mathbf{v}(\mathbf{r})|$ is the fluid speed at
$\mathbf{r}$, and $v(\mathbf{r}) = 0$ inside the solid phase. The index `S' at $T_n^\mathrm{S}$
indicates that this quantity is to be calculated on a surface (cross-section), whereas `V' at $T_n^\mathrm{V}$
indicates a volumetric quantity.

Using a simple capillary model Koponen \textit{et al.}\
concluded that $T_n^\mathrm{S} = T_n^\mathrm{V}$, but it is not difficult to show that this
is not true in a general case. To this end it suffices to consider a bundle of straight and wavy
cylindrical capillaries of the same radius and different lengths.
The contribution of each of such capillaries to the integrals in
(\ref{eq:Koponen-tortuosity-integralS}) is proportional to the area of its cross-section with $A$,
which readily implies that $T_n^\mathrm{S}$ is $A$-dependent.
For this reason $T_n^\mathrm{S}$  should not be used to calculate the hydraulic tortuosity unless,
perhaps, in very large, homogeneous systems where the effects of $A$-dependence could be averaged out.

In their actual calculations Koponen \textit{et al.}\
used Eq.~(\ref{eq:Koponen-tortuosity-integralV})  with the integrals approximated by sums \cite{Koponen96},
\begin{equation}
  \label{eq:Koponen-tortuosity-sum}
    (T_n^\mathrm{V})^n
      \approx
    \frac{\sum_i \widetilde\lambda^n(\mathbf{r}_i) v(\mathbf{r}_i)}{\sum_{i} v(\mathbf{r}_i)},
\end{equation}
where $\mathbf{r}_i$ are some points that sample uniformly the available pore space, either
by being chosen at random \cite{Koponen96} or
by being identified with the nodes of a lattice used to model the system \cite{Koponen97}.
While for a given steady velocity field $T_n^\mathrm{V}$
is a mathematically well-defined quantity to which the r.h.s.\ of (\ref{eq:Koponen-tortuosity-sum})
should converge as the number of points $\mathbf{r}_i$ goes to infinity,
the fact that it is defined through volumetric integrals introduces some
additional, presumably unintentional weighting that favors longer streamlines.
Using a method described in Sec.~\ref{sec:new_method}
it can be shown that for flows of incompressible fluids without eddies,  $T_n^\mathrm{V}$
can be expressed in terms of surface integrals
\begin{equation}
  \label{eq:Artur1}
   (T_n^\mathrm{V})^n =\frac{\int_A \widetilde\lambda^{n+1}(\mathbf{r}) v_{\perp}(\mathbf{r})   \, d^2\mathbf{r}}%
                         {\int_A \widetilde\lambda(\mathbf{r})v_{\perp}(\mathbf{r})    \, d^2\mathbf{r}},
\end{equation}
where $v_{\perp}$ is the component of the fluid velocity normal to  surface $A$.
Comparison of this formula with Eq.~(\ref{eq:Koponen-tortuosity-integralS})
confirms that  $T_n^\mathrm{V} \neq T_n^\mathrm{S}$.

Another approach was proposed by Matyka \textit{et al.}\ \cite{Matyka08}, whose formula
for the hydraulic tortuosity (here denoted by $T^\mathrm{M}$)
in an integral representation
can be written as
\begin{equation}
  \label{eq:Tortuosity-Matyka-integral}
    T^\mathrm{M} = \frac{\int_{A} \widetilde\lambda (\mathbf{r}) v_{\perp}(\mathbf{r}) \,d^2 \mathbf{r}}%
             {\int_{A}                                v_{\perp}(\mathbf{r})  \, d^2 \mathbf{r}}.
\end{equation}
where the surface $A$ need not be perpendicular to the macroscopic flow direction
and can even be curved, and $\widetilde\lambda $ and $v_{\perp}$ are assumed to vanish
inside the solid phase of the medium.
In contrast to Eqs.~(\ref{eq:Koponen-tortuosity-integralS}) and (\ref{eq:Koponen-tortuosity-integralV}),
in which tortuosities of individual streamlines were weighted with local fluid speeds,
in Eq.~(\ref{eq:Tortuosity-Matyka-integral}) they are weighted with local fluxes.
This guarantees that  for incompressible flows both integrals in
Eq.~(\ref{eq:Tortuosity-Matyka-integral}), and hence $T^\mathrm{M}$, are independent of $A$.
The actual numerical calculations were performed using a two-dimensional model system
 and Eq.~(\ref{eq:Tortuosity-Matyka-integral}) was approximated
with an arithmetic mean
\begin{equation}
 \label{eq:Matyka-sum}
   T^\mathrm{M} \approx \frac{1}{N} \sum_{i=1}^N \widetilde\lambda(\mathbf{r}_i),
\end{equation}
where $\mathbf{r}_i$ are some points on a cross-section satisfying
a constant-flux constraint between two neighboring streamlines
and $N$ is the number of these points.

Of the three methods presented in this overview, only that of Matyka \textit{et al.}\
correctly addresses the problem of recirculation zones---their contribution to
$ T^\mathrm{M}$ vanishes.
Another advantage of Eq.~(\ref{eq:Matyka-sum}) is that all terms in the sum are
of the same order of magnitude, whereas the sums in (\ref{eq:Koponen-tortuosity-sum})
contain terms that can differ by several orders of magnitude.
It is a consequence of a fact  that Eq.~(\ref{eq:Koponen-tortuosity-sum})
implicitly divides the space into regions of approximately equal volume
and assigns to them equal importance,
whereas flow in a porous medium takes place mainly in a few conducting channels
which occupy only a small fraction of the porous space.
For this reason one can expect that for the same number of streamlines
Eq.~(\ref{eq:Matyka-sum}), which assigns equal importance to equal fluid fluxes,
will be loaded with a much smaller numerical error.
A disadvantage of
Eq.~(\ref{eq:Matyka-sum}) is that it would be difficult to apply it
to three-dimensional problems, the main difficulty being to find
the points $\mathbf{r}_i$ satisfying a constant-flux condition.

Note also that it follows from Eqs.~(\ref{eq:Artur1}) and (\ref{eq:Tortuosity-Matyka-integral}) that
\begin{equation}
  \label{eq:TM=T-1V}
     T^\mathrm{M} = T_{-1}^\mathrm{V}
\end{equation}
for arbitrary incompressible flows. %This relationship has not been noticed so far.

%%%%%%%%%%%%%%%%%%%%%%%%%%%%%%%%%%%%%%%%%%%%%%%%%%%%%%%%%%%%%%%%%%%%%%%%%%%%%%%%%%%%%
%%%%%%%%%%%%%%%%%%%%%%%%%%%%%%%%%%%%%%%%%%%%%%%%%%%%%%%%%%%%%%%%%%%%%%%%%%%%%%%%%%%%%
%%%%%%%%%%%%%%%  SECTION %%%%%%%%%%%%%%%%%%%%%%%%%%%%%%%%%%%%%%%%%%%%%%%%%%%%%%%%%%%%
%%%%%%%%%%%%%%%%%%%%%%%%%%%%%%%%%%%%%%%%%%%%%%%%%%%%%%%%%%%%%%%%%%%%%%%%%%%%%%%%%%%%%
%%%%%%%%%%%%%%%%%%%%%%%%%%%%%%%%%%%%%%%%%%%%%%%%%%%%%%%%%%%%%%%%%%%%%%%%%%%%%%%%%%%%%

\section{Alternative method     \label{sec:new_method} }

Just as it is possible to express a volumetric integral (\ref{eq:Koponen-tortuosity-integralV})
as a surface integral  (\ref{eq:Artur1}), it is possible to express a surface integral
(\ref{eq:Tortuosity-Matyka-integral}) as a volumetric integral.
The resulting formula reads
\begin{equation}
  \label{eq:Artur-calkowa}
    T^\mathrm{M}  =
      \frac{\int_{V} { v(\mathbf{r}) }\, d^3\mathbf{r} }%
    {\int_{V} v_x(\mathbf{r}) \, d^3\mathbf{r}},
 \end{equation}
where $v_x$ denotes the velocity component parallel to the macroscopic flow direction.
This equation can be written in a particularly simple form
\begin{equation}
  \label{eq:Artur-iloraz}
  T^\mathrm{M} =
     \frac{\langle v \rangle }{\langle v_x \rangle},
\end{equation}
in which $\langle\ldots\rangle$ denotes a spatial average over the pore space.
This is the main result of the paper, but before we proceed to prove it, a few remarks are in order.

Fluid streamlines are determined numerically by solving---usually  thousands of times---an ordinary
differential equation of motion of a massless fluid particle in a given velocity field.
This field is obtained by some method of the computational fluid dynamics (CFD), which yields
approximate values of the fluid velocity field at limited number of discrete points (a mesh).
To compute a streamline, the velocity at an arbitrary point is thus required, and this
is calculated using some kind of extrapolation from the values at the mesh,
which introduces additional errors. For these reasons numerical determination of streamlines is
time-consuming and error-prone. This task is especially difficult at low porosities,
as in this case the conducting channels are narrow and particularly sinuous,
and under such conditions the computer-generated streamlines tend to hit the channel walls
and terminate \cite{Koponen96,Matyka08}.
The main advantage of (\ref{eq:Artur-calkowa}) is that it enables to calculate a tortuosity
without any need to determine individual streamlines. It can be thus used in very complex
geometries, e.g.\ close to the percolation threshold in 2D flows or in large-scale,
low-porosity 3D flows.

The idea that the ratio ${\langle v \rangle }/{\langle v_x \rangle}$ can be related
to the hydraulic tortuosity is not new. Carman \cite{Carman37} used it to argue that
permeability must be proportional to $T^{-2}$, cf. Eq.~(\ref{eq:Kozeny-Carman}), rather than
to $T^{-1}$, as had been earlier postulated by Karman. Koponen \textit{et al.}\ \cite{Koponen96} used this ratio
explicitly as one of several possible definitions of the hydraulic tortuosity. However,
all these attempts were based on a simple model where a porous medium is assumed to be equivalent
to a group of parallel channels and no attempt was made towards justification
of this approach in more general cases.

\subsection{Proof of equation (\protect\ref{eq:Artur-calkowa})}

\begin{figure}
\centering
\includegraphics[width=0.7\columnwidth]{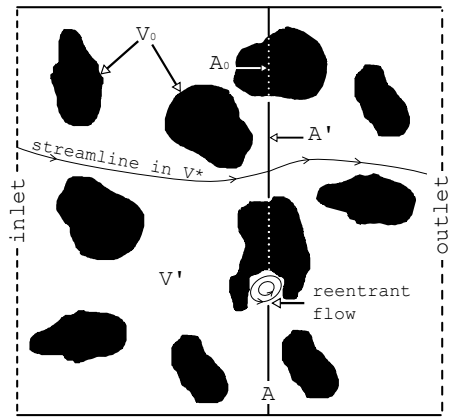}
\caption{Quantities used in the proof of equation (\protect\ref{eq:Artur-calkowa}).
Solid phase ($V_0$) is immersed in porous phase ($V'$).
A cross-section $A$ consists of the solid ($A_0$) and porous ($A'$) part.
$V^*$ is made up by all streamlines connecting the inlet and outlet planes.
Formation of eddies, e.g.\ in cavities, would violate Eq.~(\protect\ref{eq:Artur-calkowa}). \label{fig:proof}
}
\end{figure}

The system volume $V$ can be divided into two disjoint subsets, the porous space $V'$
and the solid phase $V_0$ (see Fig.~\ref{fig:proof}). Similarly, any cross-section $A$ can be divided into
$A' \equiv A \cap V'$ and  $A_0 \equiv A \cap V_0$. We
assume that $\mathbf{v} = 0$ and $\lambda =0$  at any $\mathbf{r} \in V_0$.
Let $V^* \subset V'$ denote the set of all points $\mathbf{r} \in V'$ such that
the streamline cutting $\mathbf{r}$ joins the inlet and outlet surfaces.
A flow for which $V^*\neq V'$ shall be called reentrant.

Assume that the flow is stationary, incompressible and not reentrant.
Incompressibility of flow implies that for any cross-section $A$ perpendicular to the flow direction
the denominator in Eq.~(\ref{eq:Tortuosity-Matyka-integral})
is equal to the total flux through the porous sample. This
leads to
\begin{equation}
  \int_{V} v_x(\mathbf{r}) \, d^3\mathbf{r}
    =
  L \int_{A} v_{\perp}(\mathbf{r})  \, d^2 \mathbf{r}.
\end{equation}
To prove (\ref{eq:Artur-calkowa}), it thus suffices to show that
\begin{equation}
  \label{eq:to-be-proven}
    \int_{V'} { v(\mathbf{r}) }\, d^3\mathbf{r}
      =
    \int_{A'} \lambda (\mathbf{r}) v_{\perp}(\mathbf{r}) \,d^2 \mathbf{r}.
\end{equation}

Since $\mathbf{v} (\mathbf{r})$ is defined and continuous at each $\mathbf{r}$,
and $\lambda (\mathbf{r})$ is defined and continuous at each $\mathbf{r}$
except for some points from a zero-measure subset $D\subset V'$ \cite{Matyka08},
both integrals in (\ref{eq:to-be-proven}) exist.

Let $A$ be a cross-section perpendicular to the flow direction (i.e.\ to the $x$-axis)
such that each streamline cuts $A'$ only once (e.g.\ $A$ is the inlet plane).
Let $s(\mathbf{r})$ be the distance from $A'$  to $\mathbf{r}$  along the streamline passing through $\mathbf{r}$.
Except for a zero-measure set, any point in $V'$ can
be uniquely identify by the streamline it belongs to and $s(\mathbf{r})$.
Each streamline, on the other hand, is uniquely identified by $\mathbf{r} \in A'$ belonging to this streamline.
Thus $A'$ and $s$ can be used to change the integration variables in the integral on the left-hand-side (l.h.s.\/)
of Eq~(\ref{eq:to-be-proven}) from $x,y,z$ to $s,y,z$. Since $dx/ds = v_x(\mathbf{r})/v(\mathbf{r})$,
and  $\int\!\!\int v_x\, dy dz $ is constant
along streamlines in incompressible flows, we arrive at
\begin{eqnarray}
  \label{eq:xxx}
    \int_{V'} { v(\mathbf{r}) }\, d^3\mathbf{r}
     &=&
         \int\!\!\! \int\!\!\!  \int v_x(y,z,s)\, dy dz ds \nonumber \\
     &=&
         \int_{A'} \left(  \int ds(\mathbf{r}) \right) v_x(\mathbf{r}) d^2\mathbf{r} \nonumber \\
     &=& \int_{A'} \lambda (\mathbf{r}) v_x (\mathbf{r}) \,d^2 \mathbf{r},
\end{eqnarray}
which is the right-hand-side (r.h.s.) of (\ref{eq:to-be-proven}).
Validity of (\ref{eq:to-be-proven}) can be extended to cross-sections of arbitrary shape by noticing
that $v_{\perp}(\mathbf{r}) \,d^2 \mathbf{r}$ is the flux associated
with a streamline that passes through $\mathbf{r}$ and hence is
constant along a streamline since the fluid is incompressible. For the same reason,
if a streamline cuts $A$ many times, contributions to the r.h.s.\ of (\ref{eq:to-be-proven})
from each subsequent cutting
are the same in magnitude but of alternate signs and cancel out in pairs, so that effectively
each streamline contributes to the integral as if it cut the cross-section only once.

It is interesting to notice that using the same arguments
Eq.~(\ref{eq:to-be-proven}) can be generalized to
\begin{equation}
  \label{TheThesisGeneral}
      \int_{V'} f(\mathbf{r}) v(\mathbf{r}) \, d^3\mathbf{r}
      =
    \int_{A'} f(\mathbf{r}) \lambda (\mathbf{r}) v_{\perp}(\mathbf{r}) \,d^2 \mathbf{r},
\end{equation}
where a function $f(\mathbf{r})$ has a constant value along each streamline.

\subsection{Conditions of applicability of equation (\protect\ref{eq:Artur-calkowa})}

Validity of Eq.~(\ref{eq:Artur-iloraz})  is based on two assumptions:
the fluid is incompressible and the flow is not reentrant. The latter condition is met, for example,  for
irrotational or potential flows.
This implies that Eq.~(\ref{eq:Artur-iloraz}) can be used to calculate a hydraulic tortuosity
for inviscid fluids. Another important class of potential flows are those governed by the Laplace equation,
e.g.\ tracer diffusion or electric current.  Thus we conclude that  Eq.~(\ref{eq:Artur-iloraz})
can be used to calculate diffusional or electrical tortuosities.

Real fluids, however, are viscous and as they flow through a porous medium,
some recirculation zones (eddies)
are produced by rapid changes in pore aperture or blind pore spaces. These eddies make the flow
reentrant even at very low Reynolds number.
The contribution from reentrant zones to the volumetric integral in
(\ref{eq:to-be-proven}) is strictly positive, whereas it vanishes for the surface integral.
Therefore, from a mathematical point of view, a weaker relation replaces (\ref{eq:to-be-proven}) for
general laminar viscous incompressible flows,
\begin{equation}
  \label{eq:Artur-iloraz-neq}
  T^\mathrm{M} \le
     \frac{\langle v \rangle }{\langle v_x \rangle}.
\end{equation}
However, in flows through porous porous media at low Reynolds number
the volumes where the flow is reentrant not only constitute
a small fraction of the total porous volume $V'$, but the fluid velocity in these volumes
is at least an order of magnitude smaller than that in conducting channels. Therefore
the contribution from reentrant regions to the volumetric integral in (\ref{eq:to-be-proven})
can be expected to  be negligible
and hence Eq.~(\ref{eq:Artur-iloraz}) should be also applicable to viscous
flows, at least in the low Reynolds number regime.

\section{Applications \label{sec:Applications} }

To verify usability of Eq.~(\ref{eq:Artur-iloraz}), we
employed it to find the hydraulic tortuosity in a model of
freely overlapping squares
\cite{Koponen96,Koponen97,Matyka08,Koza09}. In this model one considers a two-dimensional
lattice with a porous matrix modeled with freely overlapping
solid squares of size $a\times a$ lattice units (l.u.) placed
uniformly at random locations on a square lattice $L\times L$ l.u.\ ($1 \le a
\ll L$). The squares are fixed in space but free to overlap, and the remaining void space is filled with a
fluid to which a constant, external force is imposed along the $x$ axis to model the gravity.
This system, especially at high porosities, can be regarded as a cross-section of a fibrous
material made up of long, parallel fibers aligned perpendicularly to the flow direction.

To make sure that the percolation threshold has its usual meaning, we assumed
the system to be a rectangle of size $3L/2\times L$ with impenetrable walls
along its longer side. Obstacles of size $a\times a$ were placed only
in the central part of size $L\times L$ (see Fig.~\ref{fig:color-at-0.99}).
In this way any percolating route through the pore space was open to flow.
Measurements of all physical quantities were performed only using the data from the central, porous subsystem.
Because our system was finite, some obstacle configurations with
$\varphi \geq \varphi_\mathrm{c}$, especially close to $\varphi_\mathrm{c}$,
turned out to block the flow completely.
We rejected such configurations.
We solved the flow equations numerically in the creeping flow regime
using the Palabos (Parallel Lattice Boltzmann Solver) software
\cite{website:palabos} for $a=10$ and $L=1000$
($\varphi_\mathrm{c} \approx 0.367$ \cite{Matyka09}).

\begin{figure}
\centering
\includegraphics[width=0.95\columnwidth]{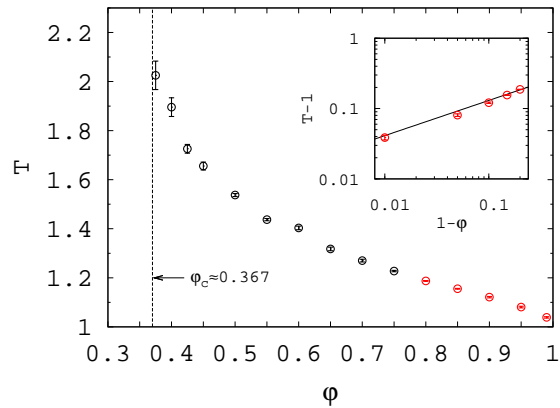}
\caption{
  (Color) Tortuosity ($T$) as a function of porosity ($\varphi$)
  in the model of overlapping squares for $a=10$ l.u.\ and $L=1000$ l.u.
  Symbols represent the results obtained with Eq.~(\protect\ref{eq:Artur-calkowa}).
  The vertical dashed line shows the percolation threshold
  $\varphi_\mathrm{c}\approx 0.367$ below which no flow
  (in the limit of $L\to\infty$) is possible.
  Inset: a log-log plot of $T-1$ as a function of $1-\varphi$; the solid line
  is the best-fit to $T - 1 \propto \sqrt{1-\varphi}$ for $\varphi\geq 0.8$.
    \label{fig:tortuosity}
}
\end{figure}

Figure \ref{fig:tortuosity}
shows the tortuosity calculated from Eq.~(\protect\ref{eq:Artur-calkowa})
for a broad range of porosities. For $0.45 \le \varphi \le 0.9$ these results
are practically the same as those reported in \cite{Matyka08}, where much smaller systems
of size $L\times L$ with periodic boundary conditions in both directions were used
and the values of physical parameters were extrapolated from those obtained for $50 \le L \le 300$.
This is consistent with \cite{Koza09}, where it was argued that the condition for the boundary
and finite-size effects to be negligible in
this model is $L\gtrsim 400$ and $a/L \lesssim 0.01$.
What is even more remarkable, very good agreement with the results of \cite{Matyka08},
where the tortuosity was calculated from Eq.~(\ref{eq:Tortuosity-Matyka-integral})
rather than from (\ref{eq:Artur-calkowa}), indicates that the difference between the two
formulas, resulting from how they treat reentrant flows, is negligible. This validates
utilization of Eq.~(\ref{eq:Artur-calkowa}) close to $\varphi_\mathrm{c}$,
where Eq.~(\ref{eq:Tortuosity-Matyka-integral}) is numerically unstable and hence rather useless.

The inset in Fig.~\ref{fig:tortuosity} depicts a log-log plot of $T-1$
as a function of $1-\varphi$ for
large porosities ($\varphi \ge 0.8$) at which the model mimics a fibrous medium.
The data suggest that in this case $T -1 \propto (1 -\varphi)^\gamma$ with $\gamma = 1/2$,
i.e.
\begin{equation}
  \label{eq:conj-T-1}
     T  = 1 + p  \sqrt{1-\varphi},
\end{equation}
where $p$ is a constant.
This finding is at odds with most of conjectures
about the tortuosity-porosity dependence for $\varphi\approx 1$,
as a vast majority of them predicts that $\gamma = 1$.
For example, Maxwell's formula for electrical conductivity of a medium
containing a dilute suspension of small spheres \cite{Maxwell1873}
implies that the electrical tortuosity $T_\mathrm{el}$ satisfies
$T_\textrm{el} = 1 + \frac12(1-\varphi)$ as $\varphi\to 1$.
Similarly, Weissberg \cite{Weissberg63} argued that
$1-\frac{1}{2}\ln\varphi\approx1 + \frac12(1-\varphi)$ is the
lower bound for diffusional tortuosity.
As for the hydraulic tortuosity, Mauret and Renauld in their study on fibrous mats
assumed that $T = 1 -p\ln\varphi$ with some constant $p$ \cite{Mauret97}.
Other hypotheses include $T = 1 +p(1-\varphi)$ \cite{Iversen93},
$T = \sqrt{1 + p(1-\varphi)}$ \cite{Boudreau06}, and
$T\propto \varphi^p$ \cite{Archie42}
(see \cite{Ahmadi2011,Clennell97,Boudreau96} for discussion of this topic) and all imply $\gamma = 1$.
While the conjectures regarding diffusional or electrical tortuosity
in highly porous media are well-grounded,
the above-mentioned conjectures regarding hydraulic tortuosity
are founded on various \emph{ad hoc} approximations
and even speculations, e.g.\ about equivalence of the hydraulic and electrical tortuosities,
and their validity is only hypothetical.
A theoretical tortuosity-porosity relation which predicts $\gamma\neq1$
was recently proposed by Ahmadi \textit{et al.}\ \cite{Ahmadi2011}. Their formula
$$
T = \sqrt{\frac{2}{3}\frac{\varphi}{1-p(1-\varphi)^{2/3}} + \frac13}
$$
implies $\gamma = 2/3$. The same value of $\gamma$ results from a formula proposed in \cite{DuPlessis91},
$T = \varphi/[1-(1-\varphi)^\frac23]$.

Since none of the above-mentioned formulas can be fitted to our numerical results,
we verified that our data
are not loaded with finite-size errors (data not shown). Then we investigated
the flow in several highly porous systems.
Typical examples of the velocity field in such systems are visualized
in Fig.~\ref{fig:color-at-0.99} (generated for $L=4000$ l.u., $a=10$ l.u.).
As can be seen, even if obstacles
occupy only $1\%$ of the volume so that practically each of them forms a separate `island',
the flow is very sinuous, as if restricted by some kind of solid-wall channels.
These virtual channels are created by variations in local concentration of obstacles.
Since a fluid flux through a 2D channel with the no-slip boundary condition
is proportional to its width cubed, the fluid passes most easily through the
interconnected regions of low local obstacle concentration,
whereas the regions of high local obstacle concentration---even if occupied
by separate obstacles---act effectively as
almost impenetrable barriers.
This many-body effect is not present in electric, diffusional or inviscid fluid  flows \cite{Koza09}.
For this reason electrical (or diffusional)
tortuosity $T_\mathrm{el}$ at high porosities is significantly lower than the hydraulic tortuosity
 and $|dT_\mathrm{el}/d\varphi|$ remains finite as $\varphi\to 1$.

\begin{figure}
\centering
\includegraphics[clip,width=0.95\columnwidth]{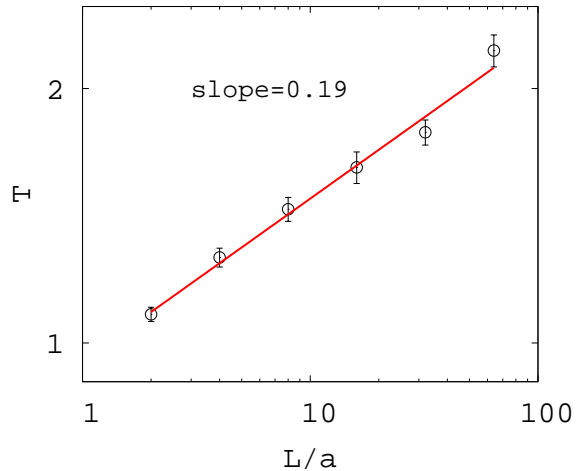}
\caption{
  (Color) Double logarithmic plot of the hydraulic tortuosity ($T$) as a function of a dimensionless system length $L/a$
  at the percolation threshold for $a=10$ l.u.\
   (symbols). A solid line is a fit to a power-law dependency $T
   \propto (L/a)^{d_\mathrm{T}}$ with ${d_\mathrm{T}} = 0.19$.
  \label{fig:T-at-prcolation}
}
\end{figure}

Figure~\ref{fig:T-at-prcolation}
presents a log-log plot  of the hydraulic tortuosity dependence
on a dimensionless system length $L/a$ at the percolation threshold  $\varphi_\mathrm{c} \approx 0.367$.
The best fit to a power law yields $T \propto L^{d_\mathrm{T}}$ with $d_\mathrm{T} = 0.19 \pm 0.01$.
This value is significantly exceeds the exponent
$d_\mathrm{min} = 1.130 \pm 0.002$  controlling the scaling of the
shortest path between two points on a percolating cluster \cite{Herrmann88}. This
indicates that even at percolation most of the fluid does not choose the shortest-path channels.
Another characteristic percolation length is the most probable traveling length
$\tilde\ell^*$, which
at bond percolation  scales  with the system size as $L^{d_{\tilde{\ell}}}$
with $d_{\tilde{\ell}} = 1.21 \pm 0.02$ \cite{Lee99}.
Using a scaling ansatz for the probability distribution function
of a path length $\lambda$ proposed in \cite{Lee99} it can be shown
that the average path length $\langle\lambda\rangle \sim \tilde\ell^* \sim L^{d_{\tilde{\ell}}}$.
Moreover, closer scrutiny of the method employed in
\cite{Lee99} to generate streamlines reveals that a constant-flux condition between
neighboring streamlines  was implicitly applied, just as in Eq.~(\ref{eq:Matyka-sum}).
Hence, $\langle\lambda\rangle/L\propto T^\mathrm{M}$, which implies
\begin{equation}
  d_\mathrm{T} =  d_{\tilde{\ell}} - 1.
\end{equation}
Our results for $d_\mathrm{T} $ are in very good agreement with this conjecture.

\section{Discussion and conclusions \label{sec:Conclusions}}

Despite of its simplicity and common usage in various areas of science,
the concept of tortuosity is poorly understood and the available literature is often misleading,
mainly because most of the theoretical research on this subject
did not go beyond much simplified  models.
Therefore in this paper we focused on the problem of calculating
the hydraulic tortuosity, defined as the average elongation of a streamline length
in a porous medium, in arbitrary flow geometries.
Our analysis shows that several existing methods of calculating hydraulic tortuosity
differ in the interpretation of
how the average streamline length is to be calculated.
Each of these methods, if applied to a system with a realistically complex geometry,
would yield a different tortuosity value, and only the method developed in \cite{Matyka08}
produces a number that does not depend on a cross-section
along which measurements are carried out and consistently addresses
the problem of recirculation zones.

For incompressible fluids the method developed in \cite{Matyka08}
can be reduced to calculating  a ratio of the mean fluid velocity
to the mean component of the fluid velocity along the
external force direction. The two methods yield exactly
the same values for regions which are connected by streamlines to the inlet and outlet surfaces,
and differ only in recirculation zones. As the contribution from recirculation zones (eddies)
is expected to be negligible at low Reynolds number regime, both methods can be considered equivalent
for incompressible creeping flows through porous media.
This conclusion was confirmed by our numerical simulations of
a 2D model of freely overlapping squares.
Thus, hydraulic tortuosity defined as the average elongation
of fluid path lengths can be calculated directly from the velocity field.
This not only greatly simplifies determination of this quantity, in experiments or numerical simulations,
including complex 3D systems,
but also opens a new perspective on its physical relevance.
Many researchers doubted if an average path length could be defined, even conceptually,
for complex flows with frequent branching and rejoining of flow streamlines
(see \cite{Clennell97}), but there is no doubt that the average fluid velocity
is a  well-defined physical quantity.
Moreover, the possibility of expressing the hydraulic tortuosity
in terms of mean fluid velocities could be used to extend its definition
to the case of higher Reynolds numbers, where the notion
of individual streamlines loses its meaning.
Note also that since no recirculation zones  can be formed in diffusional or electrical flows,
the average elongation of streamlines in diffusional or electrical
transport through porous media can  be also exactly calculated
directly from local fields, which obviates the need of determining individual streamlines.

We applied the new method in two limiting cases: very high or very low porosities.
In the former case, which corresponds to a flow through fibrous materials,
we found that the hydraulic tortuosity $T$ scales with the porosity $\varphi$
in accordance with $T-1 \propto (1-\varphi)^\gamma$, where $\gamma\approx\frac{1}{2}$.
This behavior differs from that found in diffusional or electrical flows
for which  $\gamma = 1$. This reflects a fact that determination of the velocity field
in a high-porosity hydrodynamical system is a many-body problem,
whereas the electric field in the same porous system can be safely approximated
as a superposition of single-obstacle solutions \cite{Maxwell1873,Weissberg63}.
Hydraulic and diffusional (or electrical) tortuosities are thus completely different
quantities in highly porous, fibrous systems. A difference between our result ($\gamma=1/2$)
and a recent hypothesis by Ahmadi \textit{et al.}\ \cite{Ahmadi2011}
($\gamma = 2/3$) may be caused by different
space dimensionality (2D vs 3D) and requires further investigations.

When the system is at percolation, hydraulic tortuosity was found to
scale with the system size $L$ as $L^{d_\mathrm{T}}$ with
$d_\mathrm{T} = 0.19 \pm 0.01$. This suggests that $d_\mathrm{T} = d_{\tilde{\ell}} - 1$, where
$d_{\tilde{\ell}}$ is an exponent controlling
the scaling of the most probable traveling length at bond percolation \cite{Lee99}.

\acknowledgments
This work was supported by MNiSW Grant N N519 437939.

%\bibliography{tort}

\end{document}